\pdfoutput=1
\documentclass[a4paper,12pt]{article}
\usepackage{amsmath,amssymb,amsfonts}
\usepackage{stdclsdv}
\usepackage{romannum}
\usepackage{multirow}
\usepackage{cite}

\usepackage[multiple,stable]{footmisc}
\usepackage[amsmath,amsthm,thmmarks]{ntheorem}
\allowdisplaybreaks[4]
\usepackage[noconfig]{refstyle}
\usepackage[dvipsnames]{xcolor}
\usepackage[hyperfootnotes=false, linktocpage=true, colorlinks, citecolor=blue, linkcolor=blue, urlcolor=Maroon]{hyperref}
\usepackage{array,tabularx,booktabs,geometry,subfigure,graphicx,longtable}
\usepackage[bf]{caption}
\def\bZ{\mathbb{Z}}

\def\cA{\mathcal{A}}

\def\cM{\mathcal{M}}
\def\cN{\mathcal{N}}

\usepackage{authblk}

\title{Machine Learning Free Quotients of CICYs}

\author[2]{Wei Cui \footnote{cwei@bimsa.cn}}
\author[1]{Xin Gao \footnote{xingao@scu.edu.cn}} 
\author[3,4]{Mohsen Karkheiran \footnote{mkarkhei@ualberta.ca}}
\author[2]{Juntao Wang \footnote{juntao.wang@bimsa.cn}}

\affil[1]{College of Physics, Sichuan University, Chengdu 610065, China} 
\affil[2]{Beijing Institute of Mathematical Sciences and Applications, Beijing 101408, China}
\affil[3]{Department of Mathematical and Statistical Sciences, University of Alberta, Edmonton T6G 2J5, Canada}
\affil[4]{Pure Mathematics, University of Waterloo, Waterloo N2L 3G1, Canada}

\date{}

\begin{document}
\maketitle	

\pagenumbering{arabic} 

\begin{abstract}
\noindent Free quotients of Calabi-Yau manifolds play an important role in string compactification. In this paper, we explore machine learning techniques, such as  fully connected neural networks and multi-head attention (MHA) models, as a potential approach to detect $\mathbb{Z}_2$, $\mathbb{Z}_3$, $\mathbb{Z}_4$ and $\mathbb{Z}_2\times\mathbb{Z}_2$ free quotients of CICYs. When tested on unseen examples, both models successfully identified almost all free quotients for $\mathbb{Z}_2$, $\mathbb{Z}_3$, $\mathbb{Z}_4$ and $\mathbb{Z}_2\times\mathbb{Z}_2$ symmetry. These results demonstrate that well-trained machine learning models can effectively generalize to new Calabi-Yau manifolds and may aid in the broader classification of free quotients in the future.

\end{abstract}

\thispagestyle{empty}
\setcounter{page}{0}
\newpage

\tableofcontents

\section{Introduction}
Discrete symmetries of Calabi-Yau manifolds play an essential role in string phenomenology. In heterotic string model building, for example, compactifying ten-dimensional space-time on smooth quotients of Calabi-Yau manifolds allows the use of Wilson lines to break GUT gauge group down to the Standard Model gauge group\cite{Candelas:1985en}. Meanwhile, using quotients of Calabi-Yau manifolds can give more possibilities to get exactly three families of particles as well. Beyond these smooth free quotients, string models can also be constructed on orbifolds, which include quotients of Calabi-Yau manifolds with fixed points. A notable example is provided by Type II compactifications, where orientifolds often realized as $\mathbb{Z}_2$ involution with fixed points—are widely studied (see reviews \cite{Grana:2005jc, Douglas:2006es} and references there). In general, all quotients of Calabi-Yau manifolds, whether with or without fixed points, originate from their discrete symmetries action.

Due to their significance, discrete symmetries of Calabi-Yau manifolds and their corresponding quotients have been investigated in a variety of contexts. For instance,   the orientifold Calabi-Yau threefolds with non-trivial $\mathbb{Z}_2$ divisor exchange involutions were  constructed for $h^{1,1}\leq 6$ with explicitly fixed loci and types of O-planes \cite{Gao:2013pra, Altman:2021pyc} under the fine, regular, star triangulations (FRST) from the Kreuzer-Skarke dataset of reflexive four-dimensional polyhedra \cite{Kreuzer:2000xy,Altman:2014bfa}.  The $\bZ_2$ single divisor reflection  was considered in \cite{Crino:2022zjk} for partially $h^{1,1}\leq 12$.  Recently, more than 300 million orientifolds Calabi-Yau threefolds with both multi-divisor reflections and divisor exchange involutions up to $h^{1,1} = 12$ were constructed efficiently in   \cite{Cao:2024oqx}, where new type of $\bZ_2$ free quotients were found. In \cite{Braun:2017juz}, smooth Calabi-Yau threefolds with general $\bZ_n (n\geq 3)$  free quotient symmetry  were classified for $h^{1,1}\leq 3$.

In the context of Complete Intersection Calabi-Yau 3-folds (CICYs) we focused in this paper \cite{Candelas:1987kf},    all smooth free quotient CICYs  associated with discrete symmetries in the original CICY list has been originally classified in \cite{Braun:2010vc}, where the quotient symmetry actions were descended from linear actions on the homogeneous coordinates of the ambient projective spaces. 
In \cite{Braun:2010vc}, the author introduced a character-valued index method that replaces the cumbersome fixed-point calculations with purely algebraic ones, avoiding primary decomposition and elimination.   This approach was later applied in \cite{Gray:2021kax} to identify cyclic free quotients on the most favorable description of CICYs \cite{Anderson:2017aux}.  The Hodge number for those free quotient CICYs classified in \cite{Braun:2010vc} were determined in  ~\cite{Candelas:2015amz, Constantin:2016xlj}. A landscape of $\bZ_2$  orientifold CICYs with fixed locus together with free quotient CICYs  were constructed in \cite{Carta:2020ohw} based on the  favorable  CICYs database \cite{Anderson:2017aux}.

Although significant systematic progress has been made in studying discrete symmetries of Calabi-Yau manifolds, 
traditional methods for finding them, such as checking fixed loci or using character-valued indices, are either slow or computationally expensive, especially for large datasets. The remaining challenges are twofold. First, a large portion of Calabi-Yau manifolds remains unexplored. For toric cases, systemically searches for $\bZ_2$ free quotient orientifolds have so far been limited  to $h^{1,1}\leq 12$ \cite{Cao:2024oqx} while general  $\bZ_n (n\geq 3)$ free quotients have only been studied for $h^{1,1}\leq 3$ \cite{Braun:2017juz}. For CICYs, existing classifications of symmetries depend on the specific form of the configuration matrix, and new symmetries continue to be discovered even for different representations of the same CICY \cite{Gray:2021kax}. This means that to find all smooth free quotients of a CICY, one may need to consider all its different representations, which can make the number of cases extremely large. So the systematic study of CICY quotients remains largely untouched beyond the original CICYs \cite{Candelas:1987kf} and its favorable representation \cite{Anderson:2017aux}. 
Second, there is still no general and efficient method to determine whether a quotient is free. For example, the character-valued index method \cite{Braun:2010vc} which speeds up fixed-point freeness checks does not applied directly to toric cases. Moreover, smoothness checks often require primary decomposition and elimination, which can already be extremely time-consuming for a single Calabi-Yau manifold.

As a consequence of these challenges, machine learning offers a promising alternative. Its application to the study of the Calabi–Yau landscape is particularly natural due to the large and complex datasets involved. Machine learning has been applied in various contexts, including learning polytope structures—most relevant to this work, in identifying polytopes that give rise to orientifold Calabi–Yau hypersurfaces with non-trivial $\bZ_2$ divisor exchange involutions \cite{Gao:2021xbs}, analyzing triangulations and fibrations \cite{Altman:2018zlc,Demirtas:2020dbm,Bao:2021ofk, Berman:2021mcw,Berglund:2024reu},  computing Hodge numbers \cite{Bull:2018uow,He:2018jtw,He:2020lbz,Erbin:2020tks,Keita:2024skh,He:2024coy}, evaluating numerical metrics \cite{Anderson:2020hux,Jejjala:2020wcc,Douglas:2020hpv,Larfors:2021pbb,Berglund:2022gvm} and calculating line bundle cohomologies \cite{Klaewer:2018sfl,Brodie:2019dfx}.  Besides, machine learning has also been applied to search for some new construction of Calabi-Yau manifolds such as the so-called gCICYs \cite{Anderson:2015iia,Cui:2022cxe} and higher dimension of Calabi-Yau n-folds \cite{Aggarwal:2023swe, Alawadhi:2023gxa, Hirst:2023kdl, Keita:2025dnj}. 

Given the need for a fast method to determine fixed-point freeness and the existing examples on CICYs, 
the natural purpose of this paper is to extend the border of applying machine learning to identify free quotients of CICYs with respect to certain discrete symmetries. 
Specifically, we focus on $\bZ_2$, $\bZ_3$, $\bZ_4$ and $\bZ_2 \times \bZ_2$  symmetries, for which there is sufficient number of samples to train models. For each symmetry, we implemented both standard forward-feed, fully connected neural network and multi-head attention (MHA) models  with transformer structure \cite{DBLP:journals/corr/VaswaniSPUJGKP17}, which is extensively used in Large Language Model (LLM) such as Gemini \cite{gemini,gemini-2},  Claude \cite{claude}, DeepSeek \cite{deepseek,deepseek-2} and chat-GPT \cite{chatgpt, chatgpt-2}. Evaluating these models on CICYs not seen during training, we successfully identify all the free quotients for $\bZ_2$, $\bZ_3$, $\bZ_4$ and $\bZ_2 \times \bZ_2$ by MHA model while for fully connected model  only 3 out of 57 $\bZ_2 \times \bZ_2$ cases missed. Considering the rarity of free quotients, these results are highly encouraging. They strongly suggest that machine learning could become a powerful tool for large-scale searches of Calabi–Yau free quotients in future studies.

This paper is organized as follows: in section \ref{sec:freequotient}, we review the construction of free quotients of  CICYs, introducing the notation, explaining in detail how a quotient with respect to a given discrete symmetry is defined, and describing the traditional method for finding free quotients. In  section \ref{sec:ML} we will show how the data is prepared, details of the machine learning models and the  results. Finally in Section \ref{sec:conclusion} we will discuss more about the result and possible future directions following this work. 

\section{Free quotients of CICYs}
\label{sec:freequotient}

\subsection{Discrete symmetry in CICYs}
A complete intersection Calabi-Yau (CICY) threefold  is usually represented by  the following configuration matrix,
\begin{eqnarray} \label{config}
X=\left [ \begin{array}{c|cccc}
\mathbb{P}^{n_1} & q^1_1 & q^1_2 & \dots & q^1_k  \\
\mathbb{P}^{n_2} & q^2_1 & q^2_2 & \dots & q^2_k\\
\vdots & \vdots & \vdots & \ddots & \vdots  \\
\mathbb{P}^{n_m} & q^{m}_1 & q^{m}_2 & \dots & q^{m}_k
\end{array}
\right ]
\end{eqnarray}
In the configuration matrix, the leftmost column specifies the ambient space $\cA$ of the CICY, given by $m$ direct product of projective spaces $\cA = \mathbb{P}^{n_1}\times\mathbb{P}^{n_2}\times\dots\times\mathbb{P}^{n_m}$. Each subsequent column corresponds to a homogeneous defining polynomial in the ambient space, where the integer $q^i_j$ denotes the degree of the $j$-th polynomial along the $i$-th projective factor. These $q^i_j$  should satisfy the Calabi-Yau condition $n_i+1 = \sum^k_j  q^j_i$  for $i=1, \cdots  m$. If $X$ is a three-fold, then there should be one more condition that  $ \sum^m_i n_i=k+3$.  

All Calabi–Yau manifolds of this type with positive integer entries $q^i_j$ were simply connected, and fully classified in \cite{Candelas:1987kf}, resulting in a list of $7890$ configuration matrices, commonly referred to as the original CICY list. For any CICY with positive entries, there exists a unique matrix in this list representing it.  Since the discrete symmetries depend on the configuration matrix used to describe a CICY, it is important to emphasize their favorable descriptions \cite{Anderson:2017aux}, where the divisor classes of the manifold $X$  are  all descended from its  ambiant space $\cA$.  Within the original CICY list, $4896$ out of $7890$ configuration matrices are obviously favorable. Furthermore, \cite{Anderson:2017aux} showed through the so-called ineffective splitting, apart from $48$ exceptional CICYs, every other CICY admits at least one favorable description.

In general, determining all discrete symmetries of a given CICY remains an open problem. However, those symmetries that descend from actions on the ambient space can be well defined and fully classified once the configuration matrix of the CICY is specified. Following the terminology used in \cite{Braun:2010vc,Gray:2021kax}, such symmetries can be described using a CICY group, defined as a quadruple $(C, G, \pi_r, \pi_c)$, together with its $\pi$-representation $(G', \pi, \vec{D}, \gamma)$. The components of the quadruple $(C, G, \pi_r, \pi_c)$ are defined as follows:
\begin{itemize}
\item $C$ is the combination of information encoding the data  of the CICY configuration matrix: 
\[
(d_i, c_{ij}, \delta_{j})_{1\leq i\leq m, 1\leq j\leq k},
\]
where $m$ is the number of projective factors in the ambient space and $k$ is the number of defining polynomials of the CICY.  
$d_i$ corresponds to the dimension of the $i^{\rm th}$ projective space. 
$c_{ij}$ is the degree of homogeneity of the $j^{\rm th}$ distinct defining polynomial in the $i^{\rm th}$ projective ambient space.
Finally, $\delta_j$ specifies the multiplicity, i.e., how many times the $j$-th distinct defining polynomial’s multi-degree appears in the configuration matrix.
\item $G$ is the finte group which will act on CICY.
\item $\pi_r$ is a group homomorphism which tells us how the finte group $G$ permutes rows of configuration matrix. Formally, it could be written as 
\[
\pi_r:G\rightarrow P_{\text{row}}
\]
where $P_{row}$ a group of permutations on elements of $\vec{d}$ with the $d_i$ as entries.
\item $\pi_c$ is a group homomorphism which tells us how $G$ permutes the distinct columns of the configuration matrix. It can be formally written as
\[
\pi_c:G\rightarrow P_{\text{col}}
\]
with $P_{\text{col}}$ group of permutations on elements of $\vec{\delta}$ with $\delta_j$ as entries.
\end{itemize}
A quadruple is defined to be a CICY group of a given configuration matrix if the condition
\[
c_{ij}=c_{\pi_r(i)\pi_c(j)}
\]
is satisfied for all $i$ and $j$. In essence, the task reduces to identifying all possible combinations of row and column permutations that leave the configuration matrix invariant.

Determining the CICY group $(C, G, \pi_r, \pi_c)$ alone is not sufficient. One must also specify how the group $G$ acts on the homogeneous coordinates and on the defining polynomials. This additional information is captured by the $\pi$-representation of the CICY group, $(G^{\prime},\pi,\vec{D},\gamma)$ where
\begin{itemize}
\item  $G^{\prime}$ is a finite group which is the generalized Schur cover of group $G$\cite{Braun:2010vc}. If $G$ is cyclic, then $G^{\prime}$ is the same as $G$.
\item $\vec{D}$ is a vector which contains either $\vec{d}$ or $\vec{\delta}$.
\item $\pi$ is a group homomorphism from $G^{\prime}$ to a permutation group of $\vec{D}$.
\item Finally for $\gamma$, it tells us how $G$ acts on elements of $\vec{D}$. Formally, for each $D_i$ of $\vec{D}$, $\gamma$ gives the map $\gamma_i:D_i\rightarrow GL(D_i,\mathbb{C})$ and it satisfies the following property:
\[
\gamma_{\pi(h)(i)}(g)\gamma_i(h)=\gamma_i(gh),\quad \forall g, h\in G.
\]
By considering all the elements of $\vec{D}$, $\gamma$ can be written in the following compact way,
\[
\gamma(g)=P(\pi(g),\vec{D})\textnormal{diag}(\gamma_1,\dots,\gamma_m)
\]
where $P(\pi(g),\vec{D})$ represents how $G$ permutes elements in $\vec{D}$.
\end{itemize}
More specifically, the $\pi$ representation will naturally split into two parts. One is the action on homogeneous coordinates $(G,\pi_r,\vec{d},\gamma)$ and another is the action on defining polynomials $(G,\pi_c,\vec{\delta},\rho)$. Combine all the  information above, a CICY group action can be formally written as the tuple,
\begin{equation} \label{eq:CICYgroup}
    (C,G,\pi_r,\gamma,\pi_c,\rho)
\end{equation}
Given a CICY group action $(C, G, \pi_r, \gamma, \pi_c, \rho)$, the quotient of a CICY is defined by the set of polynomials satisfying
\[
\rho^{-1}(g)\vec{p}(\gamma(g)\vec{x})=\vec{p}(\vec{x}),
\]
which ensures that the defining polynomials are invariant under the combined action of $G$ on both the ambient space coordinates and the defining polynomials themselves. Since these definitions are rather formal, an explicit example of symmetry actions is provided in the Appendix \ref{subsec:example} to help clarify the concepts introduced in this subsection.

In practice, the symmetry action is completely specified by the row action matrix $\gamma$ and the column action matrix $\rho$. Thus, the CICY configuration matrix $C$, together with $\gamma$ and $\rho$, provides all the information needed to describe the quotient with respect to a given symmetry. Consequently, the machine learning task considered in this paper can be formulated as a classification problem on data of the form $(C, \gamma, \rho)$.

\subsection{Procedures of looking for smooth free quotients of CICYs}
The procedures of looking for free quotients on CICYs includes the following steps,
\begin{itemize}
\item{{\bf Index Calculation}: To determine the possible symmetry orders that a given CICY can admit, one requires that the order of any discrete symmetry divides all relevant topological index of the CICY. The indices of interest here are
\begin{eqnarray}
\chi({\cal N}^k \otimes TX^l) &=& \int_X \textnormal{td}(TX) \wedge \textnormal{ch}({\cal N}^k \otimes TX^l) \\ \nonumber
\sigma({\cal N}^k \otimes TX^l) &=& \int_X L(X) \wedge \tilde{\textnormal{ch}}({\cal N}^k \otimes TX^l)
\end{eqnarray}
where ${\cal N}$ and $TX$ denote the normal and tangent bundles of the Calabi–Yau manifold respectively. Here $\mathrm{td}$ and $\mathrm{ch}$ represent the Todd class and the Chern character, $L(X)$ is the Hirzebruch $L$-polynomial, and $\tilde{\mathrm{ch}}$ denotes the Chern character with the curvature form rescaled by a factor of two.

In practice, evaluating these index is a relatively inexpensive step in the search for free quotients of CICYs. However, for CICYs with large configuration matrices, this computation can still become time-consuming.
}
\item{{\bf Quotient Action}: For each symmetry order obtained in the previous step, we consider all finite groups of that order and construct their possible actions on the CICY’s homogeneous coordinates and defining polynomials. This step is purely algebraic, and while it is generally not computationally intensive, the complexity can increase significantly for CICYs with large configuration matrices.} 
\item{{\bf Fixed Locus}: For each of the above group actions, we test whether the action introduces fixed points on the quotient.  Significant simplification can been acheived by introducing  the character-valued index
which we reviewed in Appendix \ref{subsec:index}.   To be short, we can calculate the character-valued index for each finite group element $g \in G$.
\begin{equation}
\label{eq:index}
\chi(g)(L)=\sum_i(-1)^i\textnormal{Tr}_{H^i(X,L)}(\gamma^*(g))
\end{equation}
where for each group action $\gamma(g)$, there is an induced action on an invariant line bundle $L$, $\gamma^*(g) L \longrightarrow L $, which  induce an action on the cohomology $H^i(X,L)$. 
If the symmetry action is fixed point free, then either $\chi(g) = 0, \forall g \neq 1 $ for $L$  an invariant line bundle with respect to the symmetry, or $\chi(g)(1)$ be an integer multiple of the order of the group for $L$  in addition equivariant.
Thus by computing equation (\ref{eq:index}) for judicious choices of $L$ one can rule out many possible symmetries in an efficient manner if they have fixed points. However, this step remains one of the most computationally demanding in the entire procedure. } 
\item {{\bf  Smoothness Check}: Finally, for all quotients that are free of fixed points, one must test whether they are smooth. This step is particularly time-consuming, as it requires performing primary decomposition and elimination, both of which are computationally intensive. }
\end{itemize}
Among the steps described above, the first two are relatively straightforward, while the third and fourth are computationally demanding. In this work, we focus on these steps using machine learning, for which we already have a sufficiently large set of examples. The computational challenge arises from two main factors: the number of possible group actions can be extremely large, often up to $10^{10}$ for a single manifold, and testing each action with traditional methods requires checking whether the fixed loci intersect the Calabi–Yau defining polynomials, a process involving costly primary decomposition and elimination. The character-valued index method simplifies this by replacing these computations with index calculations, but it only reduces the number of candidate free quotients, which still need verification. For larger datasets, such as toric Calabi–Yau manifolds or alternative CICY representations, the process remains slow. This underscores the value of machine learning for accelerating the last two steps, which motivates the present study.

\section{Machine learning the freely acting symmetries}
\label{sec:ML}

In this section, we will implement various machine learning models to find free quotients of CICYs with different discrete symmetry groups. We will show that both feedforward, fully connected neural network and multi-head attention models can effectively learn if a given group action is free or has fixed points on CICYs. Determining if a given symmetry $G$ is freely acting or has fixed points on CICYs can be formulated as a binary classification in machine learning. In subsection \ref{subsec:3.1}, we introduce how to prepare the data by data compression and set up the general strategy to use them in machine learning. Next, we apply concrete machine learning models to study free quotients of CICYs using those data including fully connected neural network models in subsection \ref{subsec:3.2} and multi-head attention models  in subsection \ref{subsec:3.3}.

\subsection{Data Preparation}
\label{subsec:3.1}

The input data are given in terms of CICY group defined in equation \eqref{eq:CICYgroup}. Since $(\pi_r,\pi_c)$ can be naturally induced from $(C, \gamma, \rho)$, we can simply use 
\[
(C,G,\pi_r,\gamma,\pi_c,\rho) \to (C,\gamma,\rho)
\]
which is a combination of three matrices. The output is denoted by $1 $ when $G$ is acting freely on the CICY and $0$ otherwise, i.e. 
\begin{equation}
(C,\gamma,\rho)\rightarrow 1\quad \textnormal{or} \quad  0.
\end{equation}
However, the input data may contain many redundant zeros which does not give any useful information, but takes extra memory in machine learning. To get rid of them while maintaining the essential feature of the group action, we introduce a method to compress input data from the triplet $(C,\gamma,\rho)$ into a single matrix.

\subsubsection*{Data compression}

 To illustrate the process of data compression, here we take a case from $\mathbb{Z}_2$ symmetry as an example. Cases from other symmetries are similar. The CICY configuration matrices $C$, group actions $\gamma$ and $\rho$ come with different sizes. For CICYs with $\bZ_2$ symmetry, the maximal size of configuration matrix $C$ is $12$ by $15$, with the largest size of $\rho$ followed as $15$ by $15$ and the largest size of $\gamma$ is $30$ by $30$. In preparing the data, one needs to pad all data into the following maximal size, 
\begin{equation}\label{shape}
(C_{12,15},\gamma_{30,30},\rho_{15,15})
\end{equation}
with sub-indices representing the shape of matrices. However, this is not an effective way to organize the input data. 

For example, configuration matrix $C$ of most CICYs with $\bZ_2$ symmetry have much smaller size compared with $12 \times 15$. After padding, majority entries of $C$ are zeroes that won't contribute in machine learning algorithm. Similar for the other two matrices $\gamma$ and $\rho$. This format has at least the following drawbacks:
\begin{itemize}
\item Putting input data in form of equation\eqref{shape} waste lots of memory and CPU/GPU hours.
\item  Large number of redundant zeros may weaken the desired feature and bring confusions to neural networks. 
\end{itemize} 
The format of input data can be greatly simplified by noticing that $\gamma$ and $\rho$ matrix has only one non-zero entries on each row or each column because they represent the group action on homogeneous coordinates of ambient space and defining polynomials that can be either permutation or multiplication of a phase. We will introduce an efficient way to compress the input data given in equation \eqref{shape} to a single matrix. 

It is convenient to explain the idea with the following example. Consider the CICY with configuration matrix,
\begin{equation}\label{cicyex}
C=\begin{pmatrix}
1 & 1 & 0 & 0 & 0 & 0 & 0 \\
1 & 0 & 1 & 0 & 0 & 0 & 0 \\
2 & 0 & 0 & 0 & 0 & 0 & 0 \\
0 & 0 & 0 & 1 & 1 & 0 & 1 \\
0 & 0 & 0 & 1 & 0 & 1 & 1 \\
0 & 1 & 1 & 0 & 1 & 1 & 0 \\
\end{pmatrix}.
\end{equation}
It has a $\mathbb{Z}_2$ symmetry 
with the action on homogeneous coordinates as
\begin{equation}
\gamma=
\left(
\begin{array}{*{16}{c}}
-1 & 0 & 0 & 0 & 0 & 0 & 0 & 0 & 0 & 0 & 0 & 0 & 0 & 0 & 0 & 0 \\
0 & 1 & 0 & 0 & 0 & 0 & 0 & 0 & 0 & 0 & 0 & 0 & 0 & 0 & 0 & 0 \\
0 & 0 & -1 & 0 & 0 & 0 & 0 & 0 & 0 & 0 & 0 & 0 & 0 & 0 & 0 & 0 \\
0 & 0 & 0 & 1 & 0 & 0 & 0 & 0 & 0 & 0 & 0 & 0 & 0 & 0 & 0 & 0 \\
0 & 0 & 0 & 0 & -1 & 0 & 0 & 0 & 0 & 0 & 0 & 0 & 0 & 0 & 0 & 0 \\
0 & 0 & 0 & 0 & 0 & 1 & 0 & 0 & 0 & 0 & 0 & 0 & 0 & 0 & 0 & 0 \\
0 & 0 & 0 & 0 & 0 & 0 & 0 & 0 & 0 & 1 & 0 & 0 & 0 & 0 & 0 & 0 \\
0 & 0 & 0 & 0 & 0 & 0 & 0 & 0 & 0 & 0 & 1 & 0 & 0 & 0 & 0 & 0 \\
0 & 0 & 0 & 0 & 0 & 0 & 0 & 0 & 0 & 0 & 0 & 1 & 0 & 0 & 0 & 0 \\
0 & 0 & 0 & 0 & 0 & 0 & 1 & 0 & 0 & 0 & 0 & 0 & 0 & 0 & 0 & 0 \\
0 & 0 & 0 & 0 & 0 & 0 & 0 & 1 & 0 & 0 & 0 & 0 & 0 & 0 & 0 & 0 \\
0 & 0 & 0 & 0 & 0 & 0 & 0 & 0 & 1 & 0 & 0 & 0 & 0 & 0 & 0 & 0 \\
0 & 0 & 0 & 0 & 0 & 0 & 0 & 0 & 0 & 0 & 0 & 0 & -1 & 0 & 0 & 0 \\
0 & 0 & 0 & 0 & 0 & 0 & 0 & 0 & 0 & 0 & 0 & 0 & 0 & -1 & 0 & 0 \\
0 & 0 & 0 & 0 & 0 & 0 & 0 & 0 & 0 & 0 & 0 & 0 & 0 & 0 & 1 & 0 \\
0 & 0 & 0 & 0 & 0 & 0 & 0 & 0 & 0 & 0 & 0 & 0 & 0 & 0 & 0 & 1 \\
\end{array}
\right)
\end{equation} 
and the action on defining polynomials is,
\begin{equation}
\rho=
\left(
\begin{array}{ccccccc}
1 & 0 & 0 & 0 & 0 & 0 & 0 \\
0 & 1 & 0 & 0 & 0 & 0 & 0 \\
0 & 0 & 1 & 0 & 0 & 0 & 0 \\
0 & 0 & 0 & 1 & 0 & 0 & 0 \\
0 & 0 & 0 & 0 & 0 & 1 & 0 \\
0 & 0 & 0 & 0 & 1 & 0 & 0 \\
0 & 0 & 0 & 0 & 0 & 0 & 1 \\
\end{array}
\right)
\end{equation}
This $\bZ_2$ symmetry defined by $\gamma$ and $\rho$ turns out to be a free action on the CICY with configuration matrix $C$. The data compression algorithm is as follows: 
\begin{itemize}
\item Firstly, generate a list starting from 1 to the number of ambient space coordinates (or number of defining polynomials). For the case now, from $\gamma$ one can get,
\begin{equation}
\{1,2,3,4,5,6,7,8,9,10,11,12,13,14,15,16\}
\end{equation}
and from $\rho$ one can get
\begin{equation}
\{1,2,3,4,5,6,7\}
\end{equation}
\item Secondly, for the $i$th coordinate (or defining polynomial) , if it is multiplied by a factor $\alpha$, we simply multiply $i$ in the list by $\alpha$. If the $i$th and $j$th coordinates are permuted, then we simply permute $i$ and $j$ in the list. Following this rule, for the example at hand, the simplified representation for $\gamma$ is,
\begin{equation}\label{eq:rowAct}
\{-1,2,-3,4,-5,6,10,11,12,7,8,9,-13,-14,15,16\}
\end{equation}
and that for $\rho$ is
\begin{equation} \label{eq:colmAct}
\{1,2,3,4,6,5,7\}
\end{equation}
In this way, $\gamma$ and $\rho$ could be simplified.
It is straightforward to generalize it to symmetries other than $\bZ_2$. 

\item 
Next, we pad the configuration matrix $C$ to the size of $12$ by $15$ that is the maximum size of CICYs configuration matrix studied in this work. The padding is performed by adding zeroes and un-padded configuration matrix $C$ are placed at the up-left corner. For example, equation \eqref{cicyex} can be put in the form,
\begin{equation}\label{cicyex2}
C_{12\times 15}=\left(
\begin{array}{*{15}{c}}
1 & 1 & 0 & 0 & 0 & 0 & 0 & 0 & 0 & 0 & 0 & 0 & 0 & 0 & 0\\
1 & 0 & 1 & 0 & 0 & 0 & 0 & 0 & 0 & 0 & 0 & 0 & 0 & 0 & 0\\
2 & 0 & 0 & 0 & 0 & 0 & 0 & 0 & 0 & 0 & 0 & 0 & 0 & 0 & 0\\
0 & 0 & 0 & 1 & 1 & 0 & 1 & 0 & 0 & 0 & 0 & 0 & 0 & 0 & 0\\
0 & 0 & 0 & 1 & 0 & 1 & 1 & 0 & 0 & 0 & 0 & 0 & 0 & 0 & 0\\
0 & 1 & 1 & 0 & 1 & 1 & 0 & 0 & 0 & 0 & 0 & 0 & 0 & 0 & 0\\
0 & 0 & 0 & 0 & 0 & 0 & 0 & 0 & 0 & 0 & 0 & 0 & 0 & 0 & 0\\
0 & 0 & 0 & 0 & 0 & 0 & 0 & 0 & 0 & 0 & 0 & 0 & 0 & 0 & 0\\
0 & 0 & 0 & 0 & 0 & 0 & 0 & 0 & 0 & 0 & 0 & 0 & 0 & 0 & 0\\
0 & 0 & 0 & 0 & 0 & 0 & 0 & 0 & 0 & 0 & 0 & 0 & 0 & 0 & 0\\
0 & 0 & 0 & 0 & 0 & 0 & 0 & 0 & 0 & 0 & 0 & 0 & 0 & 0 & 0\\
0 & 0 & 0 & 0 & 0 & 0 & 0 & 0 & 0 & 0 & 0 & 0 & 0 & 0 & 0\\
\end{array}
\right)
\end{equation}

\item 
Finally, we pad the row action part in equation \eqref{eq:rowAct} to a list with length $30$. Then, fold it into a $2$ by $15$ matrix and concatenate to the padded configuration matrix. Similarly, the column action part in equation \eqref{eq:colmAct} is padded to a list of length $15$ and also added to configuration matrix as the last row. For the example at hand, what we get will be,
\begin{equation}\label{cicyex1}
\cM=\left(
\begin{array}{*{15}{c}}
1 & 1 & 0 & 0 & 0 & 0 & 0 & 0 & 0 & 0 & 0 & 0 & 0 & 0 & 0\\
1 & 0 & 1 & 0 & 0 & 0 & 0 & 0 & 0 & 0 & 0 & 0 & 0 & 0 & 0\\
2 & 0 & 0 & 0 & 0 & 0 & 0 & 0 & 0 & 0 & 0 & 0 & 0 & 0 & 0\\
0 & 0 & 0 & 1 & 1 & 0 & 1 & 0 & 0 & 0 & 0 & 0 & 0 & 0 & 0\\
0 & 0 & 0 & 1 & 0 & 1 & 1 & 0 & 0 & 0 & 0 & 0 & 0 & 0 & 0\\
0 & 1 & 1 & 0 & 1 & 1 & 0 & 0 & 0 & 0 & 0 & 0 & 0 & 0 & 0\\
0 & 0 & 0 & 0 & 0 & 0 & 0 & 0 & 0 & 0 & 0 & 0 & 0 & 0 & 0\\
0 & 0 & 0 & 0 & 0 & 0 & 0 & 0 & 0 & 0 & 0 & 0 & 0 & 0 & 0\\
0 & 0 & 0 & 0 & 0 & 0 & 0 & 0 & 0 & 0 & 0 & 0 & 0 & 0 & 0\\
0 & 0 & 0 & 0 & 0 & 0 & 0 & 0 & 0 & 0 & 0 & 0 & 0 & 0 & 0\\
0 & 0 & 0 & 0 & 0 & 0 & 0 & 0 & 0 & 0 & 0 & 0 & 0 & 0 & 0\\
0 & 0 & 0 & 0 & 0 & 0 & 0 & 0 & 0 & 0 & 0 & 0 & 0 & 0 & 0\\
-1 & 2 & -3 & 4 & -5 & 6 & 10 & 11 & 12 & 7 & 8 & 9 & -13 & -14 & 15\\
16 & 0 & 0 & 0 & 0 & 0 & 0 & 0 & 0 & 0 & 0 & 0 & 0 & 0 & 0\\
1 & 2 & 3 & 4 & 6 & 5 & 7 & 0 & 0 & 0 & 0 & 0 & 0 & 0 & 0\\
\end{array}
\right)
\end{equation}

\end{itemize}
Through this method, we are able to compress the input data from a combination of $3$ matrices $(C,\gamma,\rho)$ into a single $15 \times 15$ matrix $\cM$.

\subsubsection*{Data enhancement}

In general, the success of any machine learning algorithms require large number of data. However, only a limited number of free quotients of CICYs are known, whereas it is comparatively easy to construct symmetry actions with fixed points. For instance, there are only $568$ free $\bZ_2$ actions on $166$ CICYs while millions of $\bZ_2$ actions with fixed points can be readily generated on the same set of manifolds.

To enlarge the training set of free quotients, we generate additional examples by applying random row and column permutations to the known $568$ free $\bZ_2$ cases. In doing so, both the configuration matrix and the corresponding group actions must be permuted consistently within the $15 \times 15$ matrix $\cM$. As an illustration, consider the matrix $\cM$ in equation \eqref{cicyex1}. Suppose we perform the following row and column permutation:
\begin{equation*}
\begin{aligned}
\textnormal{row action}:& \;(8, 9, 10, 12, 5, 6, 3, 11, 4, 2, 7, 1) \\
 \textnormal{column action}:& \; (7, 9, 3, 1, 4, 15, 11, 2, 5, 13, 14, 6, 10, 8, 12).
\end{aligned}
\end{equation*}
It is important to note that we permute not only the padded configuration matrix in the first $12$ rows of $\cM$, but also the last $3$ rows that encode the group actions. The permutation rule for these last three rows is as follows: for rows and columns containing nonzero entries, the corresponding group actions are permuted to their new positions, while for rows or columns consisting entirely of zeros, the associated group action is set to zero. For example, starting from the two actions described above, the corresponding permuted matrix is obtained immediately as
\begin{equation}
\cM'=\left(
\begin{array}{*{15}{c}}
0 & 0 & 0 & 0 & 0 & 0 & 0 & 0 & 0 & 0 & 0 & 0 & 0 & 0 & 0 \\
0 & 0 & 0 & 0 & 0 & 0 & 0 & 0 & 1 & 0 & 0 & 0 & 1 & 0 & 1 \\
1 & 0 & 0 & 0 & 1 & 0 & 0 & 0 & 0 & 1 & 0 & 0 & 1 & 0 & 0 \\
0 & 0 & 0 & 0 & 0 & 0 & 0 & 0 & 0 & 0 & 0 & 0 & 0 & 0 & 0 \\
0 & 0 & 0 & 0 & 0 & 0 & 0 & 0 & 0 & 0 & 0 & 0 & 0 & 0 & 0 \\
0 & 0 & 0 & 0 & 0 & 0 & 0 & 0 & 1 & 1 & 0 & 0 & 0 & 0 & 1 \\
1 & 0 & 0 & 1 & 0 & 0 & 0 & 0 & 0 & 0 & 0 & 0 & 0 & 0 & 0 \\
0 & 0 & 0 & 1 & 1 & 0 & 0 & 0 & 0 & 0 & 0 & 0 & 0 & 0 & 0 \\
0 & 0 & 0 & 0 & 0 & 0 & 0 & 0 & 0 & 0 & 0 & 0 & 0 & 0 & 0 \\
0 & 0 & 0 & 0 & 0 & 0 & 0 & 0 & 0 & 0 & 0 & 0 & 0 & 0 & 0 \\
0 & 0 & 0 & 2 & 0 & 0 & 0 & 0 & 0 & 0 & 0 & 0 & 0 & 0 & 0 \\
0 & 0 & 0 & 0 & 0 & 0 & 0 & 0 & 0 & 0 & 0 & 0 & 0 & 0 & 0 \\
0 & 11 & 12 & 13 & -5 & -6 & 7 & 8 & 0 & 0 & 2 & 3 & 4 & -14 & 15 \\
-16 & 17 & 0 & 0 & -20 & 21 & 0 & 0 & 0 & 0 & 0 & 0 & 0 & 0 & 0 \\
1 & 0 & 0 & 4 & 5 & 0 & 0 & 0 & 9 & 13 & 0 & 0 & 10 & 0 & 15 \\
\end{array}
\right)
\end{equation}
Following this rule, random row and column permutations allow us to generate many equivalent input matrices from a single $\cM$. These matrices provide sufficient data for training and also supply the basis for a voting mechanism to be introduced later. Although the discussion above was illustrated with a $\mathbb{Z}_2$ symmetry, the same data compression and augmentation procedure applies to CICYs with arbitrary cyclic symmetries. The generalization to $\mathbb{Z}_2 \times \mathbb{Z}_2$ is straightforward and will be presented subsequently.

\subsubsection*{Dataset generation}

We will study symmetries of CICYs described by abelian groups in particular those with one generator like $G=\bZ_2,\bZ_3,\bZ_4$ and $\bZ_2 \times \bZ_2$ with two generators. 
For a given set of CICYs with symmetry $G$, we will use supervised learning to find free quotients from them. The first step is to generate a proper dataset. For each symmetry group $G$, 
\begin{itemize}
    \item  Collect the free quotients of CICYs from \cite{Braun:2010vc} and \cite{Gray:2021kax}. Compressing them from CICY group to a single matrix. Let's denote a set of them as $\{\cM^1_i(G)\}$. 
    \item Generate a set of quotients that have fixed points on CICYs and transform them in the compressed format denoted as $\{\cM^0_i(G)\}$. 
    \item Since there are much less free quotients than those with fixed points, we perform data enhancement by random row and column permutations to increase the number of $\{\cM^1_i(G)\}$ until they have number of $\{\cM^0_i(G)\}$. 
    \item Label the data as 
    \begin{equation}
\{\cM^1_i(G)\} \to 1,\qquad   \{\cM^0_i(G)\} \to 0    
    \end{equation}
\end{itemize}
Combining them together $\{\cM_i(G)\} = \{\cM^1_i(G)\}  \cup \{\cM^0_i(G)\}$, we have obtained a balanced dataset that is important for binary classification.

To test our trained model, we will split the whole dataset into training, validation and testing subsets 
\begin{equation*}
    \{\cM_i(G)\} \to (\{\cM_j(G,\text{train})\},\{\cM_k(G,\text{valid})\},\{\cM_l(G,\text{test})\})
\end{equation*} 
The goal of spiting is to make sure that these three subsets don't contain the same CICY. Recall $\cM$ is made from configuration matrix and group action. There could be many different group actions on the same CICY. 
To satisfy this condition, we will perform data splitting in the following way. For each group $G$, 
\begin{itemize}
\item We group all CICYs with free $G$-actions according to the size of configuration matrix $C$. 
\item For $\cM$'s with the same size of $C$, i.e. CICYs defined by same sized configuration matrix, we will split them together with their $G$-action evenly into the training, validation and testing set.  

\item If there are less than $3$ CICYs defined by a given size of $C$, we add them to training set, then validation dataset. 
\end{itemize}
In this way, we split the dataset into training, validation and testing subset that does not contain the same CICY.

\subsubsection*{Learning strategy}

\begin{figure}[ht!]
    \centering
    \includegraphics[width=0.8\linewidth]{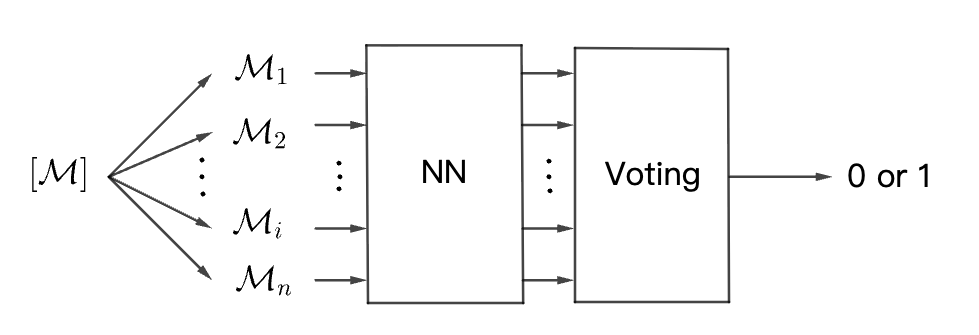}
    \caption{Voting for free quotients of CICYs with threshold $50\%$. Here $[\cM]$ as an equivalent class of data while $\cM_1$,$\cM_2$,... are different representatives.  }
    \label{fig:vot}
\end{figure}

With the dataset prepared, we proceed to train machine learning models to identify free quotients of CICYs. Following the standard pipeline, we train the models and subsequently validate and test them on the corresponding datasets. To more reliably assess model performance on previously unseen CICYs, we introduce a voting procedure, illustrated in Fig.~\ref{fig:vot}. Recall that from a given sample $\cM$, we generate $n$ equivalent representations $\cM_1,\cM_2,\ldots,\cM_n$ through random row and column permutations. The trained model is then evaluated on all $n$ versions, and the final prediction for $\cM$ is determined by majority vote, with a $50\%$ threshold. In this work, we present results both from standard supervised learning and from the enhanced voting-based approach. Remarkably, the latter enables the models to correctly identify nearly all free quotients.

\subsection{Free quotients from fully connected neural network} \label{subsec:3.2}

We will study the free quotients of CICYs using neural network made of fully connected layers. First, we will explain how to apply it to classify $\bZ_2$ quotients, followed by reporting our results for $\bZ_3$ and $\bZ_4$ quotients. Finally, we will study  $\bZ_2 \times \bZ_2$ quotients where the group has two generators.

\subsubsection*{$\mathbb{Z}_2$ symmetry}
There are $166$ out of $7,890$ CICYs admitting freely acting $\mathbb{Z}_2$ symmetry. From them, one can find $568$ free $\bZ_2$ quotients \cite{Braun:2010vc}. Besides those, there are $101$ more when favorable CICYs are considered \cite{Gray:2021kax}. To have more data for machine learning, we perform data enhancement on these known $669$ free $\bZ_2$ quotients by random row and column permutations that leads to $368,000$ data and we label them as $1$. On the same set of CICYs, we also generate $368,000$ $\bZ_2$ quotients that have fixed points \footnote{Such examples can also be found in the classification of orientifold in favorable CICYs \cite{Carta:2020ohw}.} and label them as $0$. Then, we combine these data and compress them as $15 \times 15$ matrix denoted by $\{\cM_i({\bZ_2})\}$. Following the method introduced in Subsection \ref{subsec:3.1}, we split them into $326,400$ for training, $246,400$ for validation and $163,200$ for testing.

To classify the free quotients from the data $\{\cM_i({\bZ_2})\}$, we build a neutral network with fully connected layers, where the input is a $15 \times 15$ matrix while the output is a length-$2$ vector representing the probabilities for the matrix to be $\bZ_2$ free quotient or not. The architecture is summarized in:
\begin{equation}
\label{eq:Z2}
    \cN_{225 \times 2} =(F \circ G_{64 \times 128}, F \circ G_{128 \times 256}, F \circ G_{256 \times 128}, F \circ G_{128 \times 64}, F \circ G_{64 \times 32}, S \circ G_{32 \times 2})
\end{equation}
where $G_{n\times m}$ is a fully connected layer of neurons with $n$ inputs and $m$ outputs, $F$ is the standard Relu function, BatchNorm normalization and dropout procedure,   $S$ is a softmax layer transforming 2 real numbers into a probability distribution of 2 possible outcomes.

\begin{figure}[htbp]
    \centering 
    \includegraphics[width=0.5\textwidth]{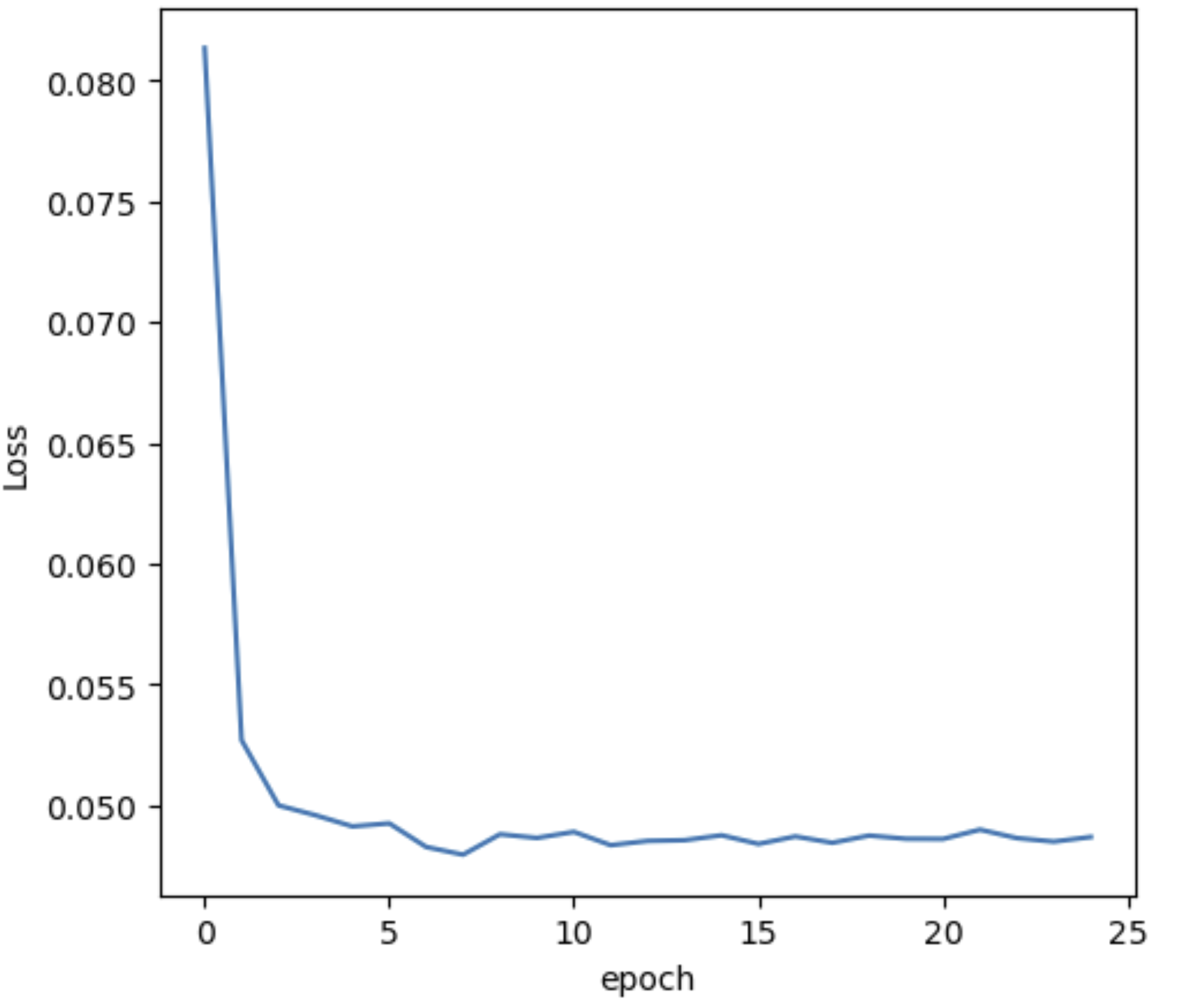}
    \caption{$\mathbb{Z}_2$ training curve for model with fully connected layers} 
    \label{z2trainingcurve} 
\end{figure}

\begin{table}[!ht]
\centering
\caption{Model performance with fully connected neural network}
\label{tab:metrics}
\begin{tabular}{lccccc}
\toprule
\textbf{Metric} & $\bZ_2$& $\bZ_3$& $\bZ_4$  & $\bZ_2 \times \bZ_2$\\
\midrule
Test Accuracy     & 0.9549 & 0.9350 & 0.9390 & 0.9433  \\
Voting & 0.9833 & 1.0 & 1.0 &0.9736\\
\bottomrule
\end{tabular}
\end{table}

We train the neutral network using $326,400$ data. Loss function is binary cross-entropy and optimizer is chosen to be Adam. The learning rate is scheduled dynamically during training. As shown in Figure \ref{z2trainingcurve}, the model converges after $10$ epochs with more than $95\%$ accuracy on both training and validation dataset. The trained model can achieve $95.49\%$ accuracy when evaluated on testing dataset as showed in table \ref{tab:metrics}, and the confusion matrix is in table \ref{tab:confusion_matrix}.

The evaluation on the original samples is given by using the voting procedure as shown in Fig.\ref{fig:vot}. For each of the equivalent classes, we generate $400$ representatives by row and column permutations. By evaluating the trained model on all these representatives and applying the voting procedure for the output with threshold $50\%$, we find that trained model can correctly predict all free $\bZ_2$-quotients and 
$145$ out of $150$ quotients that have fixed points. Thus, testing accuracy increases significantly after performing voting procedure.

\begin{table}[ht!]
\centering
\caption{Confusion matrix with fully connected neural network}
\label{tab:confusion_matrix}
\begin{tabular}{|c|cc|cc|cc|cc|}
\hline
    & \multicolumn{2}{c|}{$\bZ_2$}           & \multicolumn{2}{c|}{$\bZ_3$}           & \multicolumn{2}{c|}{$\bZ_4$}           & \multicolumn{2}{c|}{$\bZ_2\times\bZ_2$} \\ \hline
A/P     & \multicolumn{1}{c|}{Class 0} & Class 1 & \multicolumn{1}{c|}{Class 0} & Class 1 & \multicolumn{1}{c|}{Class 0} & Class 1 & \multicolumn{1}{c|}{Class 0}  & Class 1 \\ \hline
Class 0 & \multicolumn{1}{c|}{75,608}  & 5,992   & \multicolumn{1}{c|}{4,176}   & 624     & \multicolumn{1}{c|}{4,405}   & 595     & \multicolumn{1}{c|}{22,103}   & 697     \\ \hline
Class 1 & \multicolumn{1}{c|}{1,369}   & 80,231  & \multicolumn{1}{c|}{0}       & 4,800   & \multicolumn{1}{c|}{15}      & 4,985   & \multicolumn{1}{c|}{1,888}    & 20,912  \\ \hline
\end{tabular}
\end{table}

\subsubsection*{$\mathbb{Z}_3$ symmetry }

There are $497$ CICYs that could have $\mathbb{Z}_3$ symmetry, but only $31$ of them admit $\bZ_3$ free action. More free quotients are generated by random row and column permutations. On the same set of CICYs, we also generate equal number of $\bZ_3$ quotients with fixed points. Following the procedure explained in Subsection \ref{subsec:3.1}, we prepare $92,800$ data for training, $35,200$ ones for validation and $9,600 $ ones for testing.

We design a similar neutral network with architecture summarized in equation (\ref{eq:Z3}):
\begin{equation}
\label{eq:Z3}
    \cN_{450 \times 2} =(F \circ G_{64 \times 128}, F \circ G_{128 \times 64}, F \circ G_{64 \times 32}, S \circ G_{32 \times 2})
\end{equation}
where $G_{n\times m}$, $F$ and $S$ represent the same information as equation (\ref{eq:Z2}).
Note here the input is $15 \times 30 = 450$ because there are complex numbers in the action matrices, we need to split them into real and imaginary parts. Other hyperparameter are chosen to be the same as $\bZ_2$ case. The model converge fast during training and lead to $93.50\%$ accuracy on testing dataset. Other metric and confusion matrix are listed in table \ref{tab:metrics} and \ref{tab:confusion_matrix}. After the voting procedure, the trained model can correctly classify all $\bZ_3$-quotients.

\subsubsection*{$\mathbb{Z}_4$ symmetry }

We can find $667$ CICYs that allow an order $4$ symmetry, but only $23$ of them have $\mathbb{Z}_4$ free quotients. Following the same procedure like $\bZ_3$ case, we prepare training, validation and testing dataset which contain $185,000$, $45,000$ and $10,000$ samples. The architecture and choices of hyperparameter in the neural network are almost the same as the $\bZ_3$ case and we won't repeat it here. The model after training gives $93.90\%$ accuracy on testing dataset. After the voting procedure, the trained model can correctly predict all the free quotients as well as quotients with fixed points. More metric and confusion matrix are shown in table \ref{tab:metrics} and \ref{tab:confusion_matrix}.

\subsubsection*{$\mathbb{Z}_2\times\mathbb{Z}_2$ symmetry}

Different from cyclic groups, $\mathbb{Z}_2\times\mathbb{Z}_2$ symmetry has two generators. To specify group action on configuration matrix of CICY $C$, one needs two pairs of matrix $(\gamma,\rho)_i$ with $i=1,2$, one for each generator. The simplified CICY group is thus given by the following five matrices 
\begin{equation} \label{eq:z2z25}
\{C, (\gamma_1,\rho_1), (\gamma_2,\rho_2)\}.
\end{equation}
Similar as data compression for cyclic group discussed in Subsection \ref{subsec:3.1}, one can compress the above data into a single matrix as
\begin{itemize}
\item Pad configuration matrix with zeros to size $12$ by $15$.
\item For each $\mathbb{Z}_2$ factor, transform row action $\gamma_i$ into a matrix of size $2 \times 15$, and column action $\rho_i$ into a matrix of size $1 \times 15$. Concatenate them to form a matrix of size $3 \times 15$.
\item Attach two $3 \times 15$ matrix, one for each $\bZ_2$ factor, to the padded configuration matrix $C$.
\end{itemize}
Through this procedure, we compress the data given in equation \eqref{eq:z2z25} into a single $18 \times 15$ denoted by $\cM$.

There are $40$ free $\mathbb{Z}_2\times\mathbb{Z}_2$  quotients from $667$ CICYs which allow an order $4$ symmetry. Similar to cyclic group, we generate more free quotients by using row and column permutations on $\cM$. On the same set of CICYs, we also construct  $\mathbb{Z}_2\times\mathbb{Z}_2$  quotients containing fixed points. 
By the same procedure in Subsection \ref{subsec:3.1}, we split these data into training, validation and testing dataset. 
We prepared a training dataset with $140000$ samples. Using these training data, we trained a model with architecture summarized in equation (\ref{eq:z2z2}).
\begin{equation}
\label{eq:z2z2}
    \cN_{270 \times 2} =(F \circ G_{64 \times 128}, F \circ G_{128 \times 256}, F \circ G_{256 \times 128}, F \circ G_{128 \times 64}, F \circ G_{64 \times 32}, S \circ G_{32 \times 2})
\end{equation}
where $G_{n\times m}$, $F$ and $S$ represent the same information as equation (\ref{eq:Z2}).
The trained model can achieve a $94.33\%$ accuracy on the testing dataset with $45600$ samples. More metric and confusion matrix are listed in table \ref{tab:metrics} and \ref{tab:confusion_matrix}. For each of equivalent class $[\cM]$, we perform voting using the trained model among different representatives. With $50\%$ threshold, it correctly predict all $\mathbb{Z}_2\times\mathbb{Z}_2$ quotients with fixed points and $54$ out of $57$ for free quotients. The accuracy after voting is enhanced to $97.36\%$.

\subsection{Free quotients from multi-head attention models}
\label{subsec:3.3}
It can be seen from the work of last subsection, fully connected neural network models can indeed detect free quotients in a high accuracy, only failed on a few cases for symmetry $\mathbb{Z}_2\times\mathbb{Z}_2$. In real practice, it is natural to try different methods and do cross checks. Here in our case, we want to apply another framework which can naturally see more structures of our data. From the construction of our data, one can see easily there are correlations between the last several rows, which represent the row and column permutation of CICY, and the other rows and columns which the permutations act on. So naturally there is a correlation, and the final purpose is to show if this correlation represents a free quotient or not. In our work here, we tried to use a classifier based on self attention, which is good at detect self-correlations between input itself. And in fact, this classifier can indeed detect all the free quotients. In the rest of this section, we will first summarize the main structure of our model and then show the results and their analysis.
\subsubsection{Structure of the model}
Before introducing the detailed structure of the model we use, let't first give a short review on what self-attention is and how it can capture the long correlation of an input sequence. In a simple situation, given an input sequence $\mathbf{X} \in \mathbb{R}^{n \times d}$, where $n$ is the sequence length and $d$ is the feature dimension, the input is linearly projected into three matrices: queries ($\mathbf{Q}$), keys ($\mathbf{K}$), and values ($\mathbf{V}$):
\begin{equation}
\mathbf{Q} = \mathbf{X} \mathbf{W}_Q, \quad
\mathbf{K} = \mathbf{X} \mathbf{W}_K, \quad
\mathbf{V} = \mathbf{X} \mathbf{W}_V,
\end{equation}
where $\mathbf{W}_Q, \mathbf{W}_K \in \mathbb{R}^{d \times d_k}$ and $\mathbf{W}_V \in \mathbb{R}^{d \times d_v}$ are learned weight parameters of the model. By using $\mathbf{Q}$,$\mathbf{K}$ and $\mathbf{V}$, the output is just the weighted sum of values with attention weights derived from the compatibility between queries and keys:
\begin{equation}
\mathrm{Attention}(\mathbf{Q}, \mathbf{K}, \mathbf{V}) = \mathrm{softmax}\left( \frac{\mathbf{Q} \mathbf{K}^T}{\sqrt{d_k}} \right) \mathbf{V}.
\end{equation}
The scaling factor $\sqrt{d_k}$ plays a role of making the training process converges properly. This mechanism allows each position in the output sequence to attend to all positions in the input, effectively capturing global information of the input sequence. In practice, a simple generalization of the above mechanism, the so-called multi-head self-attention is often used to jointly attend to information from different subspaces:
\begin{equation}
\mathrm{MultiHead}(\mathbf{Q}, \mathbf{K}, \mathbf{V}) = \mathrm{Concat}(\mathrm{head}_1, \dots, \mathrm{head}_h) \mathbf{W}_O,
\end{equation}
where $\mathrm{head}_i = \mathrm{Attention}(\mathbf{X}\mathbf{W}_i^Q, \mathbf{X}\mathbf{W}_i^K, \mathbf{X}\mathbf{W}_i^V)$, and $\mathbf{W}_O \in \mathbb{R}^{h d_v \times d}$ learned output weight. This mechanism enables the model to focus on relevant parts of the input sequence adaptively, forming the foundation of powerful architectures such as the Transformer.

 Here in our paper, multi-head attention is one core part of our model. The detailed structure of the whole model is illustrated by using the $\mathbb{Z}_4$ model we used in table \ref{tab:via_model}. 
\begin{table}[!ht]
\centering
\caption{Architecture of the $\mathbb{Z}_4$ multi-head attention model}
\label{tab:via_model}
\begin{tabular}{>{\bfseries}l l l l}
\toprule
\textbf{Component} & \textbf{Layer/Operation} & \textbf{Input Shape} & \textbf{Output Shape} \\
\midrule
Input Preprocessing &
\begin{tabular}{@{}l@{}}
Normalize value: $D_{ij}/20$ \\
Normalize coords: $i/H, j/W$ \\
Concatenate: $[D_{ij}, i, j]$
\end{tabular} &
$(B, H, W)$ &
$(B, H \times W, 3)$ \\
\addlinespace

Embedding &
\begin{tabular}{@{}l@{}}
Linear(3 $\rightarrow$ 64) \\
GELU \\
Linear(64 $\rightarrow$ 64) \\
LayerNorm
\end{tabular} &
$(B, H \times W, 3)$ &
$(B, H \times W, 64)$ \\
\addlinespace

[CLS] Token &
Expand and concatenate &
$(B, H \times W, 64)$ &
$(B, 1 + H \times W, 64)$ \\
\addlinespace

Transpose &
$ \text{transpose}(0,1) $ &
$(B, 1 + H \times W, 64)$ &
$(1 + H \times W, B, 64)$ \\
\addlinespace

Self-Attention &
\begin{tabular}{@{}l@{}}
Multi-Head Attention \\
(4 heads, $d_{\text{model}}=64$) \\
Residual + Dropout (0.6) \\
LayerNorm
\end{tabular} &
$(1 + H \times W, B, 64)$ &
$(1 + H \times W, B, 64)$ \\
\addlinespace

Classification Head &
\begin{tabular}{@{}l@{}}
Take $x[0]$ ([CLS] token) \\
Linear(64 $\rightarrow$ 128) \\
BatchNorm1d(128) \\
GELU \\
Dropout(0.6) \\
Linear(128 $\rightarrow$ 2)
\end{tabular} &
$(B, 64)$ &
$(B, 2)$ \\
\bottomrule
\end{tabular}
\end{table}
As one can see from table \ref{tab:via_model}, the structure of this model is composed by the following parts,
\begin{itemize}
\item Input processing part. Here we simply extract each matrix element's value together with its position into a compact form. More specifically, 
for each batch of data sample with batch size $B$, one element $D$ of it is a matrix with dimension $H\times W$, which we take as  $15\times 15$ or $18\times 15$ in our cases. For each $D$, We put its matrix elements $D_{ij}$ together with its position into a combination of the form
\begin{equation}\label{vaandpo}
(D_{ij},i,j)
\end{equation}
\item Embedding. Now we embed matrix (\ref{vaandpo}) into a $64$ dimensional vector space and the whole process includes a linear layer which takes matrix (\ref{vaandpo}) into a $64$ dimensional vector, followed by a GELU(Gaussian Error Linear Unit) activation function, a linear layer transforming a $64$ dimensional vector into another $64$ dimensional vector and finally a normalization layer. The GELU activation function is in the form
\begin{equation}
\textnormal{GELU}(x)=x\cdot\Phi(x)
\end{equation}
where $\Phi(x)=P(X\leq x)$ with $X=\mathcal{N}(0,1)$ the umulative distribution function of the standard normal distribution. 
\item After the embedding part, we input a CLS token (Classification token) to each of our sample, which causes the change of data shape from $H\times W$ to $1+H\times W$. The CLS token here plays a role of 'messenger', it will collect the information of all elements of each matrix through the self-attention operation. Following this operation, we did a transpose for the data to make it in formate suitable for PyTorch.
\item We feed all the data to a multi-head attention block, which includes a $4$ head self-attention, and a linear layer transforming the CLS token into a $128$ dimensional vector.
\item Finally, we put the CLS token to a normal multi-layer neural network classifier. 
\end{itemize}
This is an illustrative example of how we design the basic structure of our models. In practice, we add one more multi-head attention layer for $\mathbb{Z}_2$ and $\mathbb{Z}_2\times\mathbb{Z}_2$ models, but their main structures are similar to the one shown above. 

\subsubsection{Result and Analysis}

\begin{figure}[!ht]
    \centering 
    \includegraphics[width=0.5\textwidth]{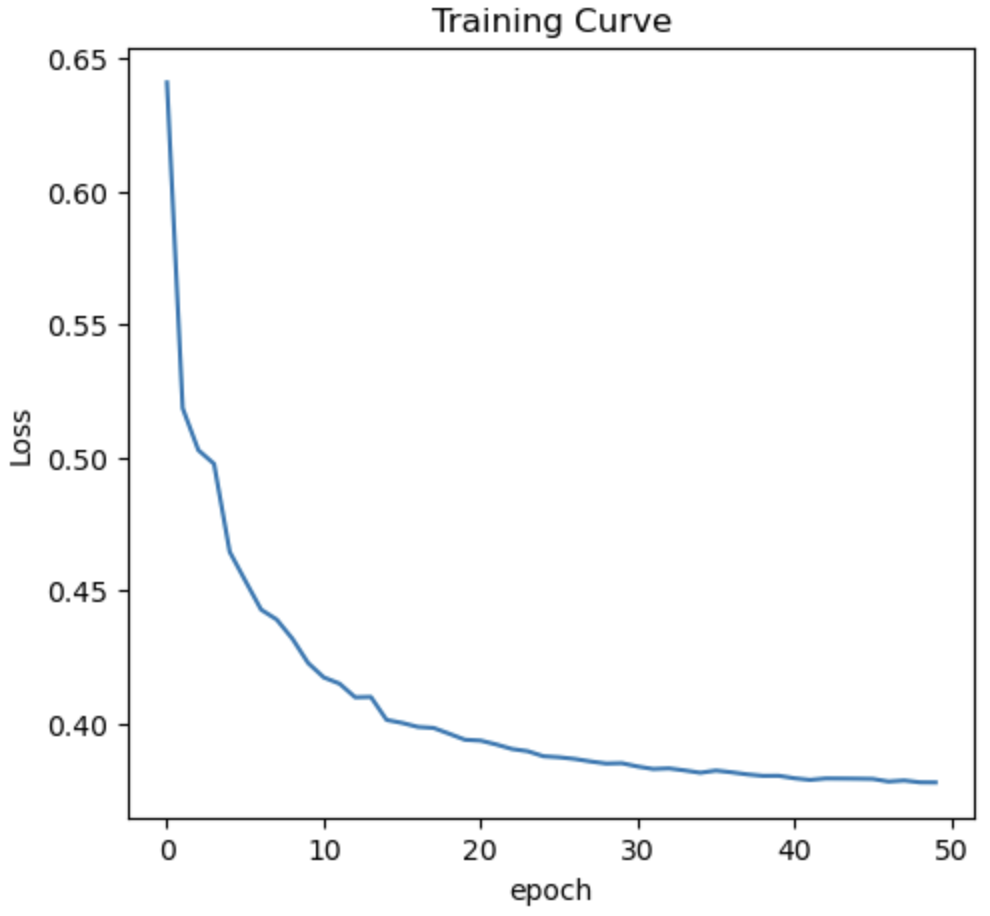}
    \caption{$\mathbb{Z}_2$ training curve for model with self-attention.} 
    \label{z2trainingcurve2} 
\end{figure}

\begin{table}[!ht]
\centering
\caption{Model performance with self-attention}
\label{tab:performance_metric2}
\begin{tabular}{lcccc}
\toprule
\textbf{Metric} & \textbf{$\mathbb{Z}_2$} & \textbf{$\mathbb{Z}_3$} & \textbf{$\mathbb{Z}_4$} & \textbf{$\mathbb{Z}_2 \times \mathbb{Z}_2$} \\
\midrule
Test Accuracy          & 0.9811 & 0.9916 &1.0 & 0.9623 \\
Voting & 0.9967 & 1.0 &1.0 & 1.0 \\
\bottomrule
\end{tabular}
\end{table}

\begin{table}[ht]
\centering
\caption{Confusion Matrices for Different Models}
\label{tab:confusion_matrix2}
\begin{tabular}{|c|c|c|c|}
\hline
\multirow{2}{*}{Model} & \multirow{2}{*}{Actual} & \multicolumn{2}{c|}{Predicted} \\
\cline{3-4}
 & & Class 0 & Class 1 \\
\hline
$\mathbb{Z}_2$ &  Class 0 & 79,261 (TN) & 2,339 (FP) \\
\cline{2-4}
 &  Class 1 & 748 (FN) & 80,852 (TP) \\
\hline
$\mathbb{Z}_3$ &  Class 0 & 4,800 (TN) & 0 (FP) \\
\cline{2-4}
 &  Class 1 & 81 (FN) & 4,719 (TP) \\
\hline
$\mathbb{Z}_4$ &  Class 0 & 4,800 (TN) & 0 (FP) \\
\cline{2-4}
 &  Class 1 & 0 (FN) & 4,800 (TP) \\
\hline
$\mathbb{Z}_2 \times \mathbb{Z}_2$ & Class 0 & 21,569 (TN) & 1,231 (FP) \\
\cline{2-4}
 & Class 1 & 486 (FN) & 22,314 (TP) \\
\hline
\end{tabular}
\end{table}

Results from the above tables are statistics on all the samples after row and column permutations. All the training process converges well and the $\mathbb{Z}_2$'s training curve is shown in Figure \ref{z2trainingcurve2}. It can be seen from table \ref{tab:performance_metric2} and \ref{tab:confusion_matrix2} that every trained model performed very well on their corresponding dataset, especially for $\mathbb{Z}_3$ and $\mathbb{Z}_4$. As for the original data, the trained model can detect all free quotients in all cases. For quotients with fixed points, $\mathbb{Z}_2\times \mathbb{Z}_2$, $\mathbb{Z}_3$ and $\mathbb{Z}_4$ models found all of them with no doubt, while the $\mathbb{Z}_2$ model failed $1$ in $150$. Compared to the fully connected neural network models, models with attention perform better not only on the prediction of the $\mathbb{Z}_3$ and $\mathbb{Z}_4$ overall data, but also on identifying more quotients with or without fixed points in the case of $\mathbb{Z}_2\times\mathbb{Z}_2$ and $\mathbb{Z}_2$. Overall, model with attention only missed one quotient with fixed points in $\mathbb{Z}_2$ case and all others are correct. This is a highly encouraging result. In addition,  this also reflects that using self-attention to grasp the global structure of the input data can indeed improve the prediction power.

\section{Conclusion and outlook}
\label{sec:conclusion}

In this work, we employed both fully connected neural networks and models with multi-head attention to detect free quotients of Calabi–Yau manifolds constructed as CICYs. The trained fully connected networks successfully predicted all $\mathbb{Z}_2$, $\mathbb{Z}_3$, and $\mathbb{Z}_4$ free quotients, failing in only $3$ out of the $57$ $\mathbb{Z}_2\times\mathbb{Z}_2$ cases. By contrast, the trained multi-head attention models achieved perfect prediction across all free quotients. Since, for a given discrete symmetry, quotients with fixed points are much more numerous than free quotients, correctly identifying the latter is of particular importance. Our results demonstrate that machine learning can serve as a practical tool to efficiently identify free quotients of CICYs or to significantly reduce the candidate set. Moreover, the models exhibit balanced performance in distinguishing quotients with and without fixed points: the neural network misclassified $5$ of the $150$ $\mathbb{Z}_2$ quotients with fixed points, while the attention model misclassified only one. For $\mathbb{Z}_3$, $\mathbb{Z}_4$, and $\mathbb{Z}_2\times\mathbb{Z}_2$, both models correctly identified all quotients with fixed points. Importantly, these results were obtained by evaluating the models on previously unseen CICYs, thereby reflecting their genuine predictive power.

In this work, the first challenge we encountered was data handling, which arises from two main issues. First, the number of available samples for training the models is limited; this can be mitigated through row and column permutations. Second, each data sample consists of 3 to 5 matrices of varying dimensions, depending on the case. To address these issues, we developed a compact representation of the row and column actions that preserves all relevant information. This representation significantly reduces the amount of data required for the models and allows all samples to be represented uniformly for symmetries with the same number of generators, which can also be applied in future studies of CICY quotients. Using this approach, we performed random row and column permutations on the CICYs and adjusted the corresponding row and column actions accordingly, generating additional training samples.

To assess the true performance of the trained models, we implemented a two-step evaluation process. First, the models were trained and evaluated on the enlarged dataset generated through row and column permutations. Second, for each original sample, we identified all corresponding permuted samples and evaluated the model’s predictions on them. If more than $50\%$ of these permuted samples were classified as free quotients, the original sample was also considered a free quotient. In general, for classification problems with limited sample sizes, such as detecting all free quotients, this kind of voting mechanism provides a robust strategy to improve prediction reliability.

In our work, we just focused on detecting free quotients of specific symmetries which have enough examples to explore.  We didn’t cover the symmetry order prediction, since it is relatively easy by using traditional method. In heterotic string compactification, we need smooth free quotients of CICYs. So after obtaining the free quotients, we also need to do the singularity check. In this current project, we didn’t cover it because there is just a few known examples which are singular but are  free quotients,  we can’t get a meaningful result from this restricted data. To check a free quotient is whether smooth or not is actually an important problem to study. When there are enough data obtained in the future, it definitely deserve more attention for its own.

Several future directions follow from this work. First, the free quotients of CICYs studied so far depend on their specific representations, and new representations may yield new free quotients. A full classification requires exploring all representations, which is extremely time-consuming using either traditional methods or character-valued indices. Machine learning offers a natural alternative for this task. Another direction is the classification of free quotients of toric Calabi-Yau manifolds, which remains incomplete. For both CICYs’ alternative representations and toric Calabi-Yau manifolds, the scarcity of known free quotients suggests that combining machine learning with character-valued indices could be particularly effective. The results of this paper highlight the potential of machine learning to advance these studies.

Additionally, for grid-shaped data, two-dimensional tensor networks such as PEPS\cite{Orus:2013kga} offer another promising approach. This method can significantly reduce problem complexity, focus on relevant regions of the parameter space, and capture non-local correlations more effectively than convolutional layers. Efficient optimization techniques, such as DMRG, are available for tensor networks, and the tensors can be reshaped into isometries, interpretable as quantum gates, allowing potential applications in quantum machine learning\cite{Rieser:2023ybj}.

\section*{Acknowledgments}

We would like to thank Xu Cao, Weicheng Xue for helpful discussions. XG is supported by National Natural Science Foundation of China under Grant No. 12375065.

\appendix
\section{An example of free quotient of CICY}
\label{subsec:example}
In this part of the appendix, we want to give an explicit example to illustrate formal concepts related to CICY group actions.
Consider CICY number 21 in the standard list, it has the following configuration matrix,
$$
\left[
\begin{array}{c|cc}
\mathbb{P}^1 & 1 & 1 \\
\mathbb{P}^1 & 0 & 2 \\
\mathbb{P}^1 & 2 & 0 \\
\mathbb{P}^1 & 0 & 2  \\
\mathbb{P}^1 & 2 & 0
\end{array}
\right].
$$
Let's define a CICY group action $(C,G,\pi_r,\gamma,\pi_c,\rho)$ for it step by step. $C$ is composed by $d$ and $\delta$. From the ambient space, one can read directly $d=(2,2,2,2,2)$, which just represents the dimension of projective spaces in the ambient space, and $\delta=(1,1)$, which represents repeatity of each column. Now we fix $C$ and turn to $G$. This CICY can have order 4 symmetries, let us take $G=\mathbb{Z}_4=(1,g,g^2,g^3)$ and take the CICY group $(C, G, \pi_r,\pi_c)$ with 
\[
\pi_r(g)=(23)(45),\quad \pi_c(g)=(12)
\]
This basically means the $\mathbb{Z}_4$ symmetry will permute the second and third row, the forth and fifth row, and at the same time permute the two columns, in this way to keep the configuration matrix invariant.

After fixing the CICY group, now we need to get the $\pi$ representation of it. Let's denote the homogeneous coordinates of the ambient space as,
\[
\left(x_{1,1},x_{1,2},x_{2,1},x_{2,2},\dots,x_{5,1},x_{5,2}\right)
\]
here the first subindex refers to which projective space the coordinate belongs to. In fact, there could be lots of $\pi$ representations of a single CICY group, here we choose the one which acts on the homogeneous coordinates in the following way,
$$
\gamma=\left(
\begin{array}{cccccccccc}
e^{i\pi/2} & 0 & 0 & 0 & 0 & 0 & 0 & 0 & 0 & 0\\
0 & -e^{i\pi/2} & 0 & 0 & 0 & 0 & 0 & 0 & 0 & 0\\
0 & 0 & 0 & 0 & 1 & 0 & 0 & 0 & 0 & 0\\
0 & 0 & 0 & 0 & 0 & -1 & 0 & 0 & 0 & 0\\
0 & 0 & 1 & 0 & 0 & 0 & 0 & 0 & 0 & 0\\
0 & 0 & 0 & 1 & 0 & 0 & 0 & 0 & 0 & 0\\
0 & 0 & 0 & 0 & 0 & 0 & 0 & 0 & 1 & 0\\
0 & 0 & 0 & 0 & 0 & 0 & 0 & 0 & 0 & -1\\
0 & 0 & 0 & 0 & 0 & 0 & 1 & 0 & 0 & 0\\
0 & 0 & 0 & 0 & 0 & 0 & 0 & 1 & 0 & 0\\
\end{array}
\right).
$$
and for the action on defining polynomials, we choose
$$
\rho=\left(
\begin{array}{cc}
0 & -1\\
1 & 0
\end{array}
\right)
$$
It is easy to check, the following defining polynomials are invariant under the action given above,
\begin{eqnarray}
&p_1=c_1x_{1,1}x_{3,1}^2x_{5,1}^2+c_2x_{1,1}x_{3,1}^2x_{5,2}^2+c_3x_{1,1}x_{5,2}x_{3,1}x_{3,2}x_{5,1}+c_4x_{1,1}x^2_{3,2}x^2_{5,1}\\\nonumber
&+c_5x_{1,1}x^2_{5,2}x^2_{3,2}+c_6x_{1,2}x_{3,1}^2x_{5,1}^2+c_7x_{1,2}x_{3,1}^2x_{5,2}^2+c_8x_{1,2}x_{5,2}x_{3,1}x_{3,2}x_{5,1}\\\nonumber
&+c_9x_{1,2}x^2_{3,2}x^2_{5,1}+c_{10}x_{1,2}x^2_{5,2}x^2_{3,2}
\end{eqnarray}

\begin{eqnarray}
&p_2=c_1x_{1,1}x_{3,1}^2x_{5,1}^2+c_2x_{1,1}x_{3,1}^2x_{5,2}^2+c_3x_{1,1}x_{5,2}x_{3,1}x_{3,2}x_{5,1}+c_4x_{1,1}x^2_{3,2}x^2_{5,1}\\\nonumber
&+c_5x_{1,1}x^2_{5,2}x^2_{3,2}+c_6x_{1,2}x_{3,1}^2x_{5,1}^2+c_7x_{1,2}x_{3,1}^2x_{5,2}^2+c_8x_{1,2}x_{5,2}x_{3,1}x_{3,2}x_{5,1}\\\nonumber
&+c_9x_{1,2}x^2_{3,2}x^2_{5,1}+c_{10}x_{1,2}x^2_{5,2}x^2_{3,2}
\end{eqnarray}
This is an example of free $\bZ_4$-quotient of CICY. 

\section{Character Valued Index}
\label{subsec:index}
We used character valued index to help us find and test all the dataset used in the current work. So in this part of appendix, we briefly summarize what character valued index is, how to calculate it and how to use it to    judge fixed point freeness. 
So what is character valued index? Just like the name itself suggested, it is certain topological index of vector bundle $V$ on a CICY $X$, but at the same time has a character vaule of certain  group $G$. As stated in the main part of this paper, for a discrete group $G$, one can form its group action   $(G,\pi_r,\gamma,\pi_c,\rho)$ on a Calabi-Yau $X$ through actions on homogeneous coordinates of ambient space $A$ of $X$. In fact, this action will naturally induce an action on cohomology group of an invariant line bundle $L$ on $X$ since the cohomology group $H^*(X,L)$ of $L$ can have a representation using homogeneous coordinates of $A$. Then the action of $G$ on the homogeneous coordinates could naturally be induced to $H^*(X,L)$. Take $\gamma(g)$ as the action of $g\in G$ and $\gamma^*(g)$ as $g$'s corresponding induced action on $H^*(X,L)$, then one can naturally define
\begin{equation}
\chi(L)(g)=\sum_i(-1)^i\textnormal{Tr}_{H^i(X,L)}(\gamma^*(g)).
\end{equation}
From this formula, one can clearly see how could a combination of character and index shows up naturally. 

Then how to calculate character valued index? Similar to the process of calculating $H^*(X,L)$, here we also need the long exact Kuszol sequence, 
\begin{equation}{\label{koszul}}
0 \to \mathcal{O}(-\sum D_j) \otimes L \to \cdots \to \left(\bigoplus_{j \leq k} \mathcal{O}(-D_j - D_k)\right) \otimes L \to \mathcal{O}(-D_i) \otimes L \to L|_X \to 0,
\end{equation}
where all the objects are vector bundles on ambient space $A$ whereas $L|_X$ a vector bundle on $X$, and $D_i$ are the divisor classes corresponding to the $i'th$ defining polynomials of CICY $X$.  In the simplest occasion, if the defining relations and thus all the maps in the Koszul sequence are $G$-invariant, then one simply obtains
\begin{equation}
\chi(L)(g) = \sum_{1<j_1<j_2<\cdots<j_p} (-1)^p \chi(L \otimes \mathcal{O}(-D_{j_1}-D_{j_2}-\cdots-D_{j_p}))|_A.
\end{equation}
for any $g$ of $G$ and $\chi(L \otimes \mathcal{O}(-D_{j_1}-D_{j_2}-\cdots-D_{j_p}))|_A$ is the character of line bundle on ambient space $A$. However as we saw in former sections, group $G$ acts non-trivially not only on defining polynomials of CICY but also on ambient space non-trivially, thus change the process of calculating $\chi(L)(g)$ in a non-trivial way. Here we refer the detailed introduction of this process to \cite{Braun:2010vc,Gray:2021kax} and just summarize the key ideas and process of calculating the character valued index. It turns out that the whole calculation process can be decomposed into a series of restriction and induction of characters according to the group actions. First of all, since group $G$ acts non-trivially on both defining polynomials and ambient space, each character $\chi(L \otimes \mathcal{O}(-D_{j_1}-D_{j_2}-\cdots-D_{j_p}))|_A$ is actually not purely a character of line bundle $L \otimes \mathcal{O}(-D_{j_1}-D_{j_2}-\cdots-D_{j_p})$ on the ambient space, but rather should be a product 
\begin{equation}{\label{singlechar}}
p_{\wedge \vec{j}}\chi(L \otimes \mathcal{O}(-D_{j_1}-D_{j_2}-\cdots-D_{j_p}))|_A.
\end{equation}
Here, $p_{\wedge \vec{j}}$ is the character taken by an orbit of polynomials which is invariant under the group action, and $\wedge \vec{j}=j_1\wedge\dots\wedge j_p$ is the representative of this orbit. $\chi(L \otimes \mathcal{O}(-D_{j_1}-D_{j_2}-\cdots-D_{j_p}))|_A$ is the character taken by the bundle $L \otimes \mathcal{O}(-D_{j_1}-D_{j_2}-\cdots-D_{j_p})$. Then the character of $\wedge \vec{j}$'s orbit will be obtained by an induction of equation (\ref{singlechar}) from group $G_{\wedge \vec{j}}$, which is the stabilizer of $G$ on $\wedge \vec{j}$, to $G$. Finally, $\chi(L)(g)$ will be an alternating sum of all the orbits which show up in equation (\ref{koszul}). Then how about the calculation of $p_{\wedge \vec{j}}$ and $\chi(L \otimes \mathcal{O}(-D_{j_1}-D_{j_2}-\cdots-D_{j_p}))|_A$? They are actually $G_{\wedge \vec{j}}$ characters which can again be calculated through induction and restriction and we refer further details to \cite{Braun:2010vc,Gray:2021kax}. Even though the process of calculating character valued index is rather involved, it turns out that it can effectively speeds up the process of getting free quotients of CICYs.

Then in practice how do we use character valued index to find free quotients? The answer of this question comes from the following important property of character valued index: if $G$-action of a CICY $X$ is fixed points free, then
\begin{itemize}
\item for a $G$-invariant line bundle $L$, $\chi(L)(g)=0$ for any $g\neq 1$ in $G$,
\item further more, for a $G$-equivariant line bundle $L$,$\chi(L)(1)$ should be divisible by the order of $G$.
\end{itemize}
By using this property, we can construct sufficient number of $G$-invariant and $G$-equivariant line bundles and calculate their character valued indices from which the number of candidates for $G$-action free quotients will be highly reduced.

\bibliographystyle{utphys}
\bibliography{bibliography}
\end{document}